\pdfoutput=1
\documentclass[%
aip,cha,
amsmath,amssymb,
reprint,%
]{revtex4-1}

\makeatletter
\def\@email#1#2{%
	\endgroup
	\patchcmd{\titleblock@produce}
	{\frontmatter@RRAPformat}
	{\frontmatter@RRAPformat{\produce@RRAP{*#1\href{mailto:#2}{#2}}}\frontmatter@RRAPformat}
	{}{}
}%
\makeatother

\usepackage{graphicx}%
\usepackage{dcolumn}%
\usepackage{bm}%
\usepackage{siunitx}%
\usepackage{nicefrac}
\usepackage{bbm} %
\usepackage[utf8]{inputenc}
\usepackage[T1]{fontenc}
\usepackage{mathptmx}
\usepackage{etoolbox}

\usepackage[pdftex]{hyperref}%
\usepackage{wasysym}

\makeatletter
\@ifclasswith{revtex4-1}{linenumbers}{\newcommand{\equationlinenobegin}{\begin{linenomath}}}{\newcommand{\equationlinenobegin}{}}
\@ifclasswith{revtex4-1}{linenumbers}{\newcommand{\equationlinenoend}{\end{linenomath}}}{\newcommand{\equationlinenoend}{}}
\makeatother

\usepackage{xr} %
\makeatletter
\newcommand*{\addFileDependency}[1]{
  \typeout{(#1)}
  \@addtofilelist{#1}
  \IfFileExists{#1}{}{\typeout{No file #1.}}
}
\makeatother

\newcommand*{\myexternaldocument}[1]{
    \externaldocument[S-]{#1}
    \addFileDependency{#1.tex}
    \addFileDependency{#1.aux}
}
\myexternaldocument{supplementary_materials2}

\newcommand{\KPZlambda}{\tilde{\lambda}}

\newcommand{\ave}[1]{\langle#1\rangle}
\newcommand{\abs}[1]{\left|#1\right|}
\newcommand{\cum}[1]{\langle#1\rangle_{\rm c}}

\newcommand{\expct}[1]{\langle#1\rangle}

\renewcommand{\eqref}[1]{Eq.~(\ref{#1})}

\newcommand{\Sk}[1]{\mathrm{Sk}[#1]}  %
\newcommand{\Ku}[1]{\mathrm{Ku}[#1]}  %
\newcommand{\largeO}[1]{\mathcal{O}(#1)}

\begin{document}
	
	\title{Initial perturbation matters: implications of geometry-dependent universal Kardar-Parisi-Zhang statistics for spatiotemporal chaos}
	
	\author{Yohsuke T. Fukai}
	\email{ysk@yfukai.net}
	\affiliation{Nonequilibrium Physics of Living Matter RIKEN Hakubi Research Team,\! RIKEN Center for Biosystems Dynamics Research,\! 2-2-3 Minatojima-minamimachi,\! Chuo-ku,\! Kobe,\! Hyogo 650-0047,\! Japan}%
	\affiliation{Department of Physics,\! The University of Tokyo,\! 7-3-1 Hongo,\! Bunkyo-ku,\! Tokyo 113-0033,\! Japan}%
	
	\author{Kazumasa A. Takeuchi}
	\affiliation{Department of Physics,\! The University of Tokyo,\! 7-3-1 Hongo,\! Bunkyo-ku,\! Tokyo 113-0033,\! Japan}%
	
	\date{\today}		
	
	\begin{abstract}
	Infinitesimal perturbations in various systems showing spatiotemporal chaos (STC) evolve following the power laws of the Kardar-Parisi-Zhang (KPZ) universality class.
	While universal properties beyond the power-law exponents, such as distributions and correlations, and their geometry dependence, are established for random growth and related KPZ systems, the validity of these findings to \textit{deterministic} chaotic perturbations is unknown.  
	Here we fill this gap between stochastic KPZ systems and deterministic STC perturbations by conducting extensive simulations of a prototypical STC system, namely the logistic coupled map lattice. 
	We show that the perturbation interfaces, defined by the logarithm of the modulus of the perturbation vector components, exhibit the universal, geometry-dependent statistical laws of the KPZ class, despite the deterministic nature of STC.
	We demonstrate that KPZ statistics for three established geometries arise for different initial profiles of the perturbation, namely point (local), uniform, and ``pseudo-stationary'' initial perturbations, the last being the statistically stationary state of KPZ interfaces given independently of the Lyapunov vector.
	This geometry dependence lasts until the KPZ correlation length becomes comparable to the system size. 
	Thereafter, perturbation vectors converge to the unique Lyapunov vector, showing characteristic meandering, coalescence, and annihilation of borders of piece-wise regions that remain different from the Lyapunov vector.
	Our work implies that the KPZ universality for stochastic systems generally characterizes deterministic STC perturbations, providing new insights for STC, such as the universal dependence on initial perturbation and beyond.
	\end{abstract}
	
	\maketitle

\begin{quotation}
Many studies have been devoted to elucidating universal properties of spatially extended dynamical systems via statistical physics. In particular, seminal work by Pikovsky, Kurths, and Politi has demonstrated that infinitesimal perturbations to various dynamical systems showing spatiotemporal chaos (STC) evolve following universal power laws of the Kardar-Parisi-Zhang (KPZ) universality class, a representative universality class for stochastic processes \cite{pikovsky_1994,pikovsky_1998}. Though exact studies on stochastic interface growth have advanced the knowledge of the KPZ class dramatically \cite{kriecherbauer_2010,corwin_2012,quastel_2015,halpin-healy_2015,sasamoto_2016,spohn_2017,takeuchi_2018}, revealing geometry-dependent but universal statistical properties beyond the power-law exponents, these developments have been attained independently from the field of dynamical systems. In this paper, we attempt to bridge those two subjects by extensive numerical simulations of the logistic coupled map lattice, a prototypical model of spatially extended dynamical systems \cite{kaneko_1989,Kaneko-1993}. We first show clear evidence that the perturbations show the universal statistical properties of the KPZ class beyond the power-law exponents, and the geometry dependence manifests itself as the dependence on the initial profile of the perturbations. We then study the time regime in which the KPZ correlation length reaches the system size and found that the perturbations converge to a unique Lyapunov vector as suggested by Oseledec's theorem \cite{Oseledec-TMMS1968,Eckmann.Ruelle-RMP1985,kuptsov_2012}. We found that this process is governed by characteristic wandering and annihilation of the boundaries of the segmental regions that differ from the Lyapunov vector. With the current knowledge on KPZ and dynamical systems, we expect our results to generally hold in STC and can be extended to general initial perturbations.
\end{quotation}

\section{Introduction}
To understand general laws 
governing spatially extended dynamical systems,
it is insightful to focus on the thermodynamic limit.
Successful approaches include 
applying the concept of phase transition 
to understand collective states  
\cite{Aranson.Kramer-RMP2002,Chate.Manneville-PRL1987,Chate.Manneville-PD1988,miller_1993,marcq_1996,*marcq_1997a} 
and synchronization \cite{Pikovsky.etal-Book2003}
in large dynamical systems.
Considering the infinite-size limit is also crucial for defining the extensivity of chaos, 
by the existence of a finite density of Lyapunov exponents per degree of freedom \cite{Ruelle-CMP1982,manneville_1985,egolf_2000}.

In particular, 
universality associated with scale invariance
has successfully provided a general insight 
for growing infinitesimal perturbations in 
spatially extended dynamical systems. 
When a system is showing spatiotemporal chaos (STC), 
i.e., chaos characterized by 
spatial and temporal disorder,
perturbations given to trajectories typically grow exponentially and fluctuate.
Pioneering studies by Pikovsky, Kurths, and Politi 
\cite{pikovsky_1994,pikovsky_1998} 
revealed
that, for a wide range of systems, 
this growing perturbation 
belongs to 
the Kardar-Parisi-Zhang (KPZ) universality class \cite{kardar_1986}, 
which is one of the most wide-ranging universality classes for 
various stochastic processes (see reviews \cite{kriecherbauer_2010,corwin_2012,quastel_2015,halpin-healy_2015,sasamoto_2016,spohn_2017,takeuchi_2018}),
including growing interfaces, %
directed polymers,
stochastic particle transport, %
nonlinear fluctuating hydrodynamics \cite{spohn_2016,spohn_2017},
and most recently, to quantum spin chains \cite{Ljubotina.etal-PRL2019,Scheie.etal-NP2021,Wei.etal-a2021,Bulchandani.etal-a2021},
to name but a few.

To be more specific, 
let us focus on a system showing STC in one-dimensional space. 
The dynamical variable is $u(x,t)$
with position $x$ and time $t$.
We consider an infinitesimal perturbation $\delta{}u(x,t)$, growing from a given initial perturbation $\delta{}u(x,0)$ applied to the trajectory.
Then, Pikovsky and coworkers \cite{pikovsky_1994,pikovsky_1998} studied fluctuations of $h(x,t):=\log\abs{\delta{}u(x,t)}$, which we call the \textit{perturbation interface}, and found universal power laws for the fluctuation amplitude $\sim{}t^{\beta}$ and the correlation length $\sim{}t^{1/z}$, with characteristic exponents $\beta=1/3$ and $z=3/2$ of the one-dimensional KPZ class.
This emergence of KPZ fluctuations in STC perturbations has been found in a variety of systems, ranging from coupled map lattices (CML) \cite{pikovsky_1994,pikovsky_1998,szendro_2007,pazo_2008,pazo_2013}, 
a coupled ordinary differential equation \cite{pazo_2008,pazo_2013}, 
partial differential equations \cite{pikovsky_1998} 
to time-delayed systems \cite{pikovsky_1998,pazo_2010a,pazo_2013}.

However, this connection to the KPZ class has far-reaching implications, beyond the exponents.
The understanding on the one-dimensional KPZ class has been completely renewed during the last decades, thanks to remarkable developments by exact solutions \cite{kriecherbauer_2010,corwin_2012,quastel_2015,halpin-healy_2015,sasamoto_2016,spohn_2017,takeuchi_2018}.
An important outcome is the determination of the distribution and correlation functions of the interface fluctuations, which turned out to be universal yet dependent on the initial condition \cite{kriecherbauer_2010,corwin_2012,quastel_2015,halpin-healy_2015,takeuchi_2018}.
More specifically, we can write down the interface height $h(x,t)$ for infinite-size systems, asymptotically, in the following form:
\equationlinenobegin\begin{equation} \label{eq:h_asymptotic}
h(x,t) \simeq v_\infty t + (\Gamma t)^{1/3} \chi(X,t),
\end{equation}\equationlinenoend
where $\chi(X,t)$ is a stochastic variable discussed below,  
$X:=x/\xi(t)$ is the coordinate rescaled by
the correlation length 
$\xi(t):=\frac{2}{A}(\Gamma{}t)^{2/3}$,
and $v_\infty,\Gamma,A$ are 
system-dependent parameters \cite{Prahofer.Spohn-PA2000,takeuchi_2018}. 
Then, it has been shown both theoretically \cite{kriecherbauer_2010,corwin_2012,takeuchi_2018} and experimentally \cite{takeuchi_2018,takeuchi_2010,takeuchi_2011,takeuchi_2012,fukai_2017,iwatsuka_2020,fukai_2020} that statistical properties of $\chi(X,t)$ are universal but depend on the choice of the initial condition $h(x,0)$.
The following constitute three canonical cases, dubbed \textit{universality subclasses}:
(i) the flat subclass for the flat initial condition $h(x,0)=0$; 
(ii) the circular subclass for wedge or curved initial conditions, e.g., $h(x,0)=-\kappa|x|$ or $-cx^2$;
(iii) the stationary subclass for statistically stationary initial conditions. 
Note that such initial conditions typically take the form of $h(x,0)=\sqrt{A}B(x)$, 
where $B(x)$ is the standard Brownian motion characterized by $\expct{[B(x+\ell)-B(x)]^2}{}={}\ell$ and $B(0)=0$. 
This is true exactly of the KPZ equation \cite{kardar_1986,barabasi_1995,takeuchi_2018} 
and also valid for other models in coarse-grained scales.
Then, for each subclass, the one-point distribution of $\chi(0,t\to\infty)$ is known to be (i) the Gaussian orthogonal ensemble (GOE) Tracy-Widom distribution \cite{tracy_1996}, (ii) the Gaussian unitary ensemble (GUE) Tracy-Widom distribution \cite{tracy_1994}, and (iii) the Baik-Rains distribution \cite{baik_2000} (see \cite{prahofer_2000,kriecherbauer_2010,corwin_2012,halpin-healy_2015,takeuchi_2018}),
denoted by the random variable $\chi_1,\chi_2,\chi_0$, respectively\footnote{
Note $\chi_1:=2^{-2/3}\chi_\text{GOE-TW}$, 
where $\chi_\text{GOE-TW}$ is the standard GOE Tracy-Widom random variable \cite{tracy_1996}.}.
These subclasses have also been characterized by correlation functions \cite{kriecherbauer_2010,corwin_2012,quastel_2015,halpin-healy_2015,sasamoto_2016,spohn_2017,takeuchi_2018}, which continue attracting considerable attention (e.g., recent developments on time covariance \cite{takeuchi_2012,ferrari_2016a,denardis_2017a,Johansson-PTRF2019,johansson_2021,liu_2021} and on general initial conditions \cite{corwin_2015,dauvergne_2019}).

With this background, the following questions arise: 
Does this initial-condition dependence of KPZ fluctuations show up in perturbation dynamics of STC?
If yes, how?
Moreover, since we expect that the vector direction of $\delta{}u(x,t)$ eventually converges to the unique Lyapunov vector associated with the largest Lyapunov exponent, as Oselsedec's theorem \cite{Oseledec-TMMS1968,Eckmann.Ruelle-RMP1985,kuptsov_2012} suggests unless the largest exponent is degenerate, how does the initial-condition dependence of $\delta{}u(x,t)$ reconcile with this fact?
In this article, we address these questions numerically 
by an extensive simulation of the CML of the logistic map, 
a prototypical model showing STC \cite{kaneko_1989,Kaneko-1993}.

\section{Method}
We employed the CML of the logistic map, defined as follows. 
The dynamical variables $u(x,t)\in[0,1]$ are defined for integer $x$ and $t$, 
with the periodic boundary condition $u(x,t)=u(\mathrm{mod}(x,L),t)$.
Time evolution is
\begin{align}
u(x,t+1)=&(1-\epsilon) f(u(x,t)) \nonumber\\
&\quad +\frac{\epsilon}{2}
[ f(u(x-1,t)) + f(u(x+1,t)) ],
\end{align}
which is denoted by $F_x(u(0,t),\dots,u(L-1,t))$,
with $f(u)=au(1-u)$ and parameters $\epsilon$ and $a$.
Here we chose $a=4$ and $\epsilon=0.05$, 
for which this CML shows fully disordered STC \cite{kaneko_1989,Kaneko-1993}.
The evolution of infinitesimal perturbations $\delta u(x,t)$ is then given by
$\delta u(x,t+1)=
\sum_{x'=0}^{L-1}\left. 
\frac{\partial F_x(u_0,\dots,u_{L-1})} {\partial u_{x'}} \right|_{u_i=u(i,t)} \delta u (x',t).$ %

To discard the transient before we applied a perturbation at $t=0$, each simulation started at $t=-T_\mathrm{warmup}$ from an initial condition $u(x,-T_\mathrm{warmup})$ generated randomly and independently.
We chose $T_\mathrm{warmup}=1000$ and checked that the system is already in the attractor at $t=0$, by comparing the histograms of $u(x,t)$ at $t=0$ and $t=10^5$ (Fig.~\ref{S-fig:z_histogram} \cite{supplementary}).
We then applied a perturbation $\delta u(x,0)$ that takes different forms as we describe below,
obtained $\delta{}u(x,t)$,
and analyzed fluctuations of $h(x,t)=\log\abs{\delta u(x,t)}$.

\begin{figure}
	\includegraphics[width=\linewidth]{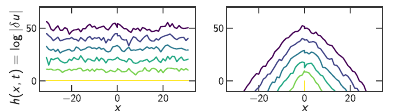}%
	\caption{\label{fig:uniform_point_snapshots} 
		Typical snapshots of $h(x,t)=\log\abs{\delta u(x,t)}$ from the uniform (left) and point (right) initial perturbations, 
		at $t=0,20,\cdots,100$ from bottom to top.
	}
\end{figure}

\section{Uniform and point initial perturbations}
We first studied the initial-shape dependence of STC perturbations.
For this, we compare the uniform initial condition $\delta u(x,0)=1$ and the point initial condition $\delta u(x,0)=\delta_{x,0}$.
Figure~\ref{fig:uniform_point_snapshots} shows typical snapshots of
the perturbation interface $h(x,t)=\log\abs{\delta{}u(x,t)}$ for those initial perturbations.
We simulated $2400$ and $160000$ independent trajectories 
with $L=2^{15}$ and $L=2^{12}$ 
for the uniform and point perturbations, respectively.

\begin{figure}
	\includegraphics[width=\linewidth]{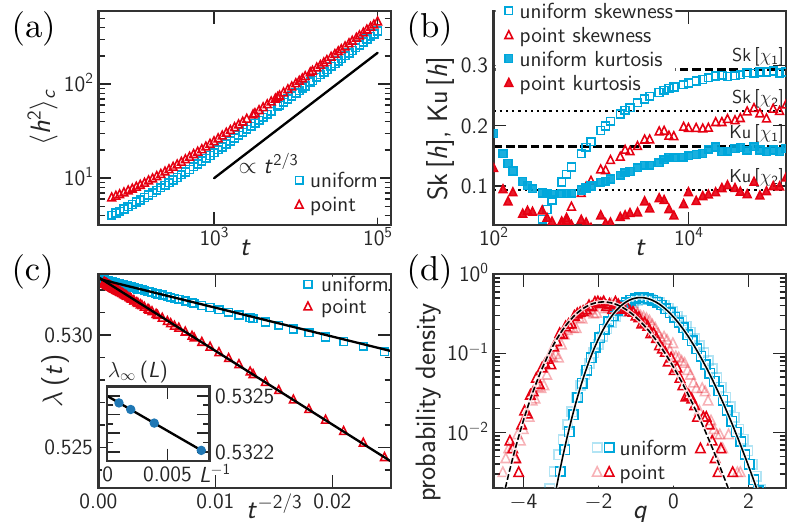}%
	\caption{\label{fig:uniform_point_results} 
		Results for the uniform and point initial perturbations.
		(a) The variance of the height $h(0,t)$. The black line is a guide to eyes indicating the power law $t^{2/3}$. 
		(b) The skewness $\Sk{h(0,t)}$ and the kurtosis $\Ku{h(0,t)}$ of the height distribution. 
		The values for $\chi_1$ and $\chi_2$ are shown by the dashed and dotted lines, respectively.
		(c) The local Lyapunov exponent $\lambda(t)$ plotted against $t^{-2/3}$. 
		The black lines are the results of the linear fit.
		(inset) The infinite-time Lyapunov exponent $\lambda_\infty(L)$ plotted against $L^{-1}$. 
		The black line is the result of the linear fit used for the estimation of $A\KPZlambda$ and $v_\infty$. 
		(d) The rescaled height distribution 
		at $t=10^3$ (lighter color) and $t=10^5$ (darker color), for the uniform (turquoise squares) and point (red triangles) perturbations.
		The probability density function of $\chi_1$ and $\chi_2$ (GOE and GUE Tracy-Widom distributions, respectively)
		is shown by the solid and dashed lines, respectively.
}
\end{figure}

First, we measured the mean height $\expct{h(0,t)}$, where $\expct{\dots}$ indicates the ensemble average. 
For the uniform perturbations, we also took the spatial average because of the translational symmetry.
As expected, $\expct{h(0,t)}$ grows linearly with time, reflecting the exponential growth of the perturbation (Fig.~\ref{S-fig:raw_cumulants} \cite{supplementary}).
We also measured the variance $\cum{h(0,t)^2}=\expct{\delta h(0,t)^2}$ with $\delta h(x,t){}:={}h(x,t)-\expct{h(x,t)}$ and the third- and fourth-order cumulants, $\cum{h(0,t)^3}=\expct{\delta h(0,t)^3}$ and $\cum{h(0,t)^4}=\expct{\delta h(0,t)^4}-3\expct{\delta h(0,t)^2}^2$.
The $k$th-order cumulants ($k=2,3,4$) show power laws $\cum{h(0,t)^k} \sim t^{k/3}$ characteristic of the KPZ class [Figs.~\ref{fig:uniform_point_results}(a) and \ref{S-fig:raw_cumulants} \cite{supplementary}], 
indicating that this exponent does not depend on the choice of initial perturbation, as expected.

We then investigated the one-point distribution of $h(0,t)$, measuring
the skewness $\Sk{h(0,t)}\allowbreak:=\allowbreak\nicefrac{\cum{h(0,t)^3}}{\cum{h(0,t)^2}^{3/2}}$ and the kurtosis  $\Ku{h(0,t)}\allowbreak:=\allowbreak\nicefrac{\cum{h(0,t)^4}}{\cum{h(0,t)^2}^{2}}$.
Figure~\ref{fig:uniform_point_results}(b) shows that those for the uniform and point initial perturbation converge to the values for the flat ($\chi_1$) and circular ($\chi_2$) KPZ subclasses, respectively.
This indicates that those perturbations indeed belong to the respective KPZ subclasses.

To characterize statistical properties further, we estimated the nonuniversal parameters, $v_\infty$, $\Gamma$, and $A$ \cite{takeuchi_2018,krug_1992}.
These parameters are expected to be independent of the initial perturbation \cite{krug_1992,takeuchi_2012,takeuchi_2018}.
First we estimated the asymptotic growth speed $v_\infty$,
which corresponds to the Lyapunov exponent in the limit $L\to\infty$.
To do so, we used the local Lyapunov exponent $\lambda(t):=\ave{\frac{\partial h(0,t)}{\partial{}t}}$, which satisfies,  for sufficiently large systems, the following relation derived from \eqref{eq:h_asymptotic}:
\begin{equation}
\lambda(t)=v_\infty+\mathrm{const.}\times t^{-2/3}.
\label{eq:v_inf_fit}
\end{equation}
Using this, we estimated $v_\infty$ by linear regression of $\lambda(t)$ against $t^{-2/3}$ [Fig.~\ref{fig:uniform_point_results}(c)].  
We thereby obtained $v_\infty=0.53249(2)$ and $0.53250(7)$ for the uniform and point initial perturbations, respectively, where the numbers in the parentheses indicate the estimation uncertainty.
This confirms that $v_\infty$ does not depend on the initial perturbation.
Next, we estimated $\Gamma$
by $\nicefrac{\cum{h(0,t)^2}}{(\cum{\chi_j^2}t^{2/3})}
\simeq\Gamma^{2/3}$ derived from \eqref{eq:h_asymptotic}, 
using $\chi_1$ and $\chi_2$ for the uniform and point initial perturbations, respectively.
Similarly to other KPZ interfaces \cite{takeuchi_2018,takeuchi_2012,ferrari_2011}, the quantity in the left-hand side shows finite-time corrections in the order of $\largeO{t^{-2/3}}$ (Fig.~\ref{S-fig:Gamma_estimation} \cite{supplementary}).
Therefore, by linear regression against $t^{-2/3}$,
we obtained $\Gamma=0.1373(5)$ and $0.1369(6)$ for the uniform and point initial perturbations, respectively.
From this, we derived the final estimate $\Gamma=0.1371(9)$.
Finally, we estimated $A$ from the asymptotic growth speed for the finite-size systems, or the Lyapunov exponent, $\lambda_\infty(L):=\lim_{t\to_\infty}\ave{\frac{\partial h}{\partial t}}$.
First, note that \eqref{eq:v_inf_fit} for $\lambda(t)$ is expected to hold until the correlation length $\xi(t)$ reaches the system size $L$. 
Beyond this, $\lambda(t)$ will deviate from \eqref{eq:v_inf_fit} and converge to $\lambda_\infty(L)\simeq v_\infty - \frac{A\KPZlambda}{L}$, where $\KPZlambda$ is a parameter satisfying $\Gamma=\frac{A^2\abs{\KPZlambda}}{2}$ \cite{krug_1992,takeuchi_2018}.
Our simulations for the uniform initial perturbation (see Table~\ref{S-tab:lambda_estimation_params} \cite{supplementary} for parameter values) confirmed that $\lambda_\infty(L)$ depends linearly on $L^{-1}$ [Fig.~\ref{fig:uniform_point_results}(c) inset].
By linear regression, we obtained $A\KPZlambda=0.0744(3)$ and $v_\infty=0.5324985(7)$, the latter refining our previous estimates.
From the estimates of $A\KPZlambda$ and $\Gamma$, we obtained $A$ by $A=\frac{2\Gamma}{A\KPZlambda}$.

With those parameters, we defined the rescaled height
\begin{align}
	q(0,t) := \frac{h(0,t)-v_\infty t}{(\Gamma t)^{1/3}} \simeq \chi(0,t) \label{eq:q_definition}
\end{align}
and investigated its statistical properties. 
Figure~\ref{fig:uniform_point_results}(d) shows 
its one-point distribution 
(again taking all $X$ for the uniform perturbation).
The distributions for the uniform and point initial perturbation 
agree with those of the exact solutions for the flat and circular subclasses, namely the GOE and GUE Tracy-Widom distributions, respectively.
Accordingly, the first- to fourth-order cumulants of $q(0,t)$ converge to those of $\chi_1$ and $\chi_2$, respectively (Fig.~\ref{S-fig:rescaled_cumulants}  \cite{supplementary}).
We also measured spatial and temporal correlation functions and found agreement with the known results for the flat and circular subclasses (see Supplemental Text \cite{supplementary}).
These results clearly indicate that the statistical properties of perturbation dynamics for the uniform and point initial perturbation 
are governed by the flat and circular KPZ subclasses, respectively.

\section{Pseudo-stationary initial perturbation}
From the viewpoint of dynamical systems theory, the perturbation interfaces are expected to converge to the unique Lyapunov vector (unique at each point in phase space), associated with the largest Lyapunov exponent \cite{Eckmann.Ruelle-RMP1985,kuptsov_2012}.
On the other hand, KPZ interfaces generally have a statistically stationary state, characterized by the time-invariance of the equal-time probability distribution of the height fluctuations.
For the KPZ equation, such a stationary state is given by the Brownian motion \cite{kardar_1986,barabasi_1995,takeuchi_2018}, $h(x,\cdot)=\sqrt{A}B(x)$, as already described.
This is characterized by the scaling law for the height-difference correlation function
\begin{align}
	C_h(\Delta x,t) := 
	\ave{[ h(x+\Delta x,t)-h(x,t)]^2}\simeq A\Delta x, 
	\label{eq:brownian_amplitude}
\end{align}
which generally holds in the stationary state of the one-dimensional KPZ class \cite{barabasi_1995,kriecherbauer_2010,corwin_2012,quastel_2015,halpin-healy_2015,sasamoto_2016,spohn_2017,takeuchi_2018}.
However, since the notion of Brownian motion can be defined only by the ensemble of realizations, for STC perturbations, it does not specify the unique realization corresponding to the Lyapunov vector.
In other words, 
all realizations of the Brownian motion except the one corresponding to the true Lyapunov vector are \textit{not} the stationary state of the perturbation interfaces.
It is therefore natural to ask whether such \textit{pseudo-stationary states}, generated randomly irrespective of the state of the dynamical system, may exhibit the characteristic fluctuation properties of the stationary KPZ subclass.

\begin{figure}
	\includegraphics[width=\linewidth]{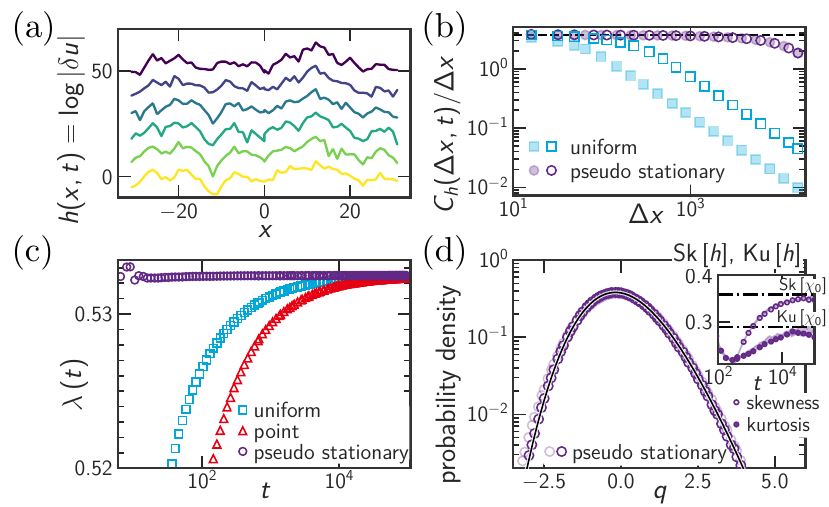}
	\caption{\label{fig:pseudo_stat_results} 
		Results for the pseudo-stationary initial perturbations.
		(a) Typical snapshots of perturbation interfaces 
		at $t=0, 20, \cdots, 100$ from bottom to top.
		(b) The height-difference correlation function $C_h(\Delta{}x,t)$ divided by $\Delta{}x$ at $t=10^4$ (lighter color) and $t=10^5$ (darker color). 
		The dashed line shows the value of $A$. 
		(c) The local Lyapunov exponent plotted against time. 
		(d) The height distribution	at $t=10^3$ (lighter color) and $t=10^5$ (darker color).
		The solid line shows the Baik-Rains distribution.
		(inset) The skewness and the kurtosis of $h(x,t)$. 
		The shaded area indicates the systematic error due to the uncertainty of $A$ in $h_0(x)=\sqrt{A}B(x)$ (see text).
		The dash-dotted lines show the values for the Baik-Rains distribution.
	}
\end{figure}

To answer this, we studied perturbation interfaces from pseudo-stationary initial conditions, 
given independently of the state of the dynamical system.
Using the estimated value of $A$, we prepared a pseudo-stationary initial perturbation
$h_0(x)=\sqrt{A}B(x)$ with $h_0(0)=0$ and the periodic boundary condition (see Supplemental Text \cite{supplementary}),
and set the initial perturbation by $\delta u(x,0)=\exp[h_0(x)]$. 
Figure~\ref{fig:pseudo_stat_results}(a) illustrates
the perturbation interface for such a pseudo-stationary initial condition.
We simulated 2400 sets of independent trajectories and pseudo-stationary initial perturbations, with $L=2^{15}$.
To evaluate systematic error due to the uncertainty of $A$ in $h_0(x)=\sqrt{A}B(x)$, we also ran simulations with the value of $A$ replaced with its upper and lower bounds of the range of uncertainty, 2400 realizations each.

We first checked the statistical stationarity of perturbation interfaces from such pseudo-stationary initial conditions. 
Figure~\ref{fig:pseudo_stat_results}(b) shows that $C_h(\Delta x,t)/\Delta x$ at $t=10^4$ and $10^5$ are identical in the pseudo-stationary case, keeping $C_h(\Delta{}x,t)/\Delta{}x\simeq{}A$ for $\Delta{}x\ll{}L$.
Figure~\ref{fig:pseudo_stat_results}(c) shows that the growth speed, or the local Lyapunov exponent, takes a constant value only after a very short period. 
These results demonstrate the statistical stationarity of our pseudo-stationary interfaces.

We then compared the statistical properties of the perturbation interfaces with the predictions for the KPZ stationary subclass.
In the following, the statistical accuracy was improved by taking advantage of translational symmetry, which guarantees that $h(x,t)-h_0(x)$ is equivalent to $h(0,t)$.
We first measured the skewness and kurtosis of $h(0,t)$ and found convergence to those of $\chi_0$, i.e., the Baik-Rains distribution characterizing the stationary subclass [Fig.~\ref{fig:pseudo_stat_results}(d) inset]. 
We also confirmed agreement with the Baik-Rains distribution, by histograms [Fig.~\ref{fig:pseudo_stat_results}(e)] and cumulants (Fig.~\ref{S-fig:rescaled_cumulants}  \cite{supplementary}) of the rescaled height $q(0,t)$.
Agreement with the stationary KPZ subclass was also confirmed through correlation functions (see Supplemental Text \cite{supplementary}), notably by quantitative agreement to the exact solution of the two-point space-time correlation \cite{prahofer_2004}.
These results demonstrate that the perturbation interfaces from the pseudo stationary initial conditions are indeed governed by the stationary KPZ subclass, despite not being stationary for the dynamical system.
Of course, one can also start with the true Lyapunov vector as an initial perturbation and the same KPZ stationary subclass is expected.

\section{Convergence to the Lyapunov vector in finite-size systems}
We have shown so far the initial-shape dependence of perturbation vectors, focusing on the three cases corresponding to the KPZ subclasses. 
On the other hand, for finite-size systems, we expect that the direction of the perturbation vectors eventually converges to that of the unique Lyapunov vector \cite{Eckmann.Ruelle-RMP1985,kuptsov_2012}, whence the initial-condition dependence is lost.
This is expected to happen when the correlation length $\xi(t)$ reaches the system size $L$.
Here we investigated how this happens for the three studied initial perturbations, using the same system but with $L=2^{10}$.
With this size, the correlation length $\xi(t)$ reaches $L$ at $t^*:=\Gamma^{-1}(\frac{AL}{2})^{3/2}\approx\num{6e5}$. 

\begin{figure}
	\includegraphics[width=\linewidth]{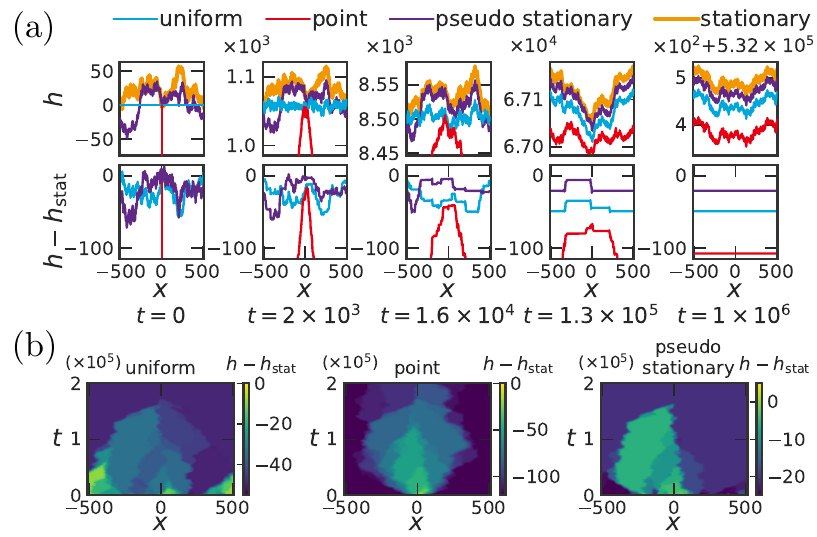}%
	\caption{\label{fig:first_LV_convergence} 
		Convergence of the perturbation interfaces to the Lyapunov vector. 
	   (a) Typical snapshots, taken at time indicated below each panel, are shown.
	   (b) Typical evolution of $h(x,t)-h_\mathrm{stat}(x,t)$.
	}
\end{figure}

To obtain the Lyapunov vector, we started simulations from $t=-T_\mathrm{warmup}-t_0$ with $t_0:=\num{6e6}(\gg{}t^*)$.
After discarding the trajectory's transient for $T_\mathrm{warmup}$ time steps, we applied a uniform perturbation $\delta{}u_0(x,-t_0)=1$ and let it evolve for $t_0$ time steps.
Then we defined $h_{\mathrm{stat}}(x,t):=\log\abs{\delta{}u_0(x,t)}-\log\abs{\delta{}u_0(0,0)}$, which is the interface corresponding to the Lyapunov vector and we call it the \textit{stationary} perturbation interface.
We compared it with the uniform, point, and pseudo-stationary interfaces $h(x,t)$ obtained for the same trajectory, and found that these indeed converge to $h_{\mathrm{stat}}(x,t)$ up to a constant shift which disappears by normalization of $\delta{}u(x,t)$ [Fig.~\ref{fig:first_LV_convergence}(a)].
Interestingly, before this convergence, the difference $h(x,t)-h_{\mathrm{stat}}(x,t)$ consists of piece-wise plateaus, reminiscent of the difference between the first and subsequent Lyapunov vectors \cite{pazo_2008}.
The plateau edges meander in an apparently random manner, coalesce, and eventually annihilate [Fig.~\ref{fig:first_LV_convergence}(b)].

\section{Summary and discussion}
In summary, using the logistic map CML as a prototypical model showing STC, we studied the distribution and correlation properties of perturbation vectors, for the following three types of initial perturbations: uniform, point, and pseudo-stationary. 
Focusing on the distribution and correlation properties, we showed clear evidence of agreement with the flat, circular, and stationary subclasses of the KPZ class, respectively.
Since the KPZ exponents have been reported for perturbations in various STC systems \cite{pikovsky_1994,pikovsky_1998,pazo_2013}, it is reasonable to believe the generality of our results.
In the pseudo-stationary case, the stationary KPZ subclass arises despite the fact that such a perturbation was not the Lyapunov vector and therefore not stationary for the dynamical system. %
These results indicate that the deterministic nature of dynamical systems does not affect the universal statistical properties of the KPZ class.
When the correlation length eventually reaches the system size, 
the perturbation interfaces from different initial perturbations 
converged to the unique stationary interface up to constant shifts, 
unlike stochastic interface growth. 
This convergence is driven by intricate meandering, coalescence, and eventual annihilation of plateau borders seen in the height difference from the stationary interface. 

Although we studied only the three types of initial perturbations here, the KPZ theory can also deal with general initial perturbations, via the \textit{KPZ variational formula} \cite{quastel_2014,corwin_2016b,quastel_2019a,fukai_2020}.
We expect that we can describe the one-point distribution of perturbation vectors for arbitrary initial conditions, by using the numerical method proposed in \cite{fukai_2020} and by estimating nonuniversal parameters according to the procedure described in this article.

Recent studies have shown that the concept of the KPZ subclasses is also useful for understanding deterministic fluctuations of STC itself, in the Kuramoto-Sivashinsky equation \cite{roy_2020,enriquerodriguez-fernandez_2021}, in anharmonic chains \cite{Mendl.Spohn-PRE2016,spohn_2016,spohn_2017}, in a discrete Gross-Pitaevskii equation \cite{Kulkarni.etal-PRA2015}, etc.
We hope that this line of research will continue providing new insights for spatially extended dynamical systems.

\section*{Supplementary Material}

See supplementary material \cite{supplementary} for Supplemental Texts on universal correlation functions and pseudo-stationary initial conditions, Table SI and Figs. S1, S2, S3, S4, S5 and S6.

\begin{acknowledgments}
We thank K. Kawaguchi and S. Ito for useful discussions on the Brownian initial condition with periodic boundary conditions. 
We are grateful to M. Pr\"{a}hofer and H. Spohn for the theoretical curves of the Baik-Rains and Tracy-Widom distributions and that of the stationary two-point correlation function, which are made available online \cite{Prahofer.Spohn-Table}, and to F. Bornemann for the curves of the Airy$_1$ and Airy$_2$ covariance, evaluated by his algorithm in \cite{Bornemann-MC2010}.
This work is supported in part by KAKENHI
 from Japan Society for the Promotion of Science
 (Grant Nos. JP17J05559, JP25103004, JP19H05800, JP20H01826).
\end{acknowledgments}

\section*{Data Availability Statement}
The data that support the findings of this study are openly available in Zenodo at https://doi.org/10.5281/zenodo.5559231.

\bibliography{citations}
\end{document}


\title{Supplemental Material for \\
		Initial perturbation matters: implications of geometry-dependent universal Kardar-Parisi-Zhang statistics for spatiotemporal chaos}%
	\author{Yohsuke T. Fukai}
	\email{ysk@yfukai.net}
	\affiliation{Nonequilibrium Physics of Living Matter RIKEN Hakubi Research Team,\! RIKEN Center for Biosystems Dynamics Research,\! 2-2-3 Minatojima-minamimachi,\! Chuo-ku,\! Kobe,\! Hyogo 650-0047,\! Japan}%
	\affiliation{Department of Physics,\! The University of Tokyo,\! 7-3-1 Hongo,\! Bunkyo-ku,\! Tokyo 113-0033,\! Japan}%
	
	\author{Kazumasa A. Takeuchi}
	\affiliation{Department of Physics,\! The University of Tokyo,\! 7-3-1 Hongo,\! Bunkyo-ku,\! Tokyo 113-0033,\! Japan}%
	\date{\today}	
	\maketitle

\onecolumngrid
\renewcommand{\thetable}{S\arabic{table}}

\section{Supplemental Text on Universal Correlation Functions}

Here we describe some notable predictions on correlation functions for the three KPZ subclasses, and the results of our data from the perturbation interfaces.
As described in the article, in the one-dimensional KPZ class, the height $h(x,t)$ is expected to take the following form for large $t$:
\begin{equation} 
h(x,t) \simeq v_\infty t + (\Gamma t)^{1/3} \chi(X,t),
\end{equation}
where $\chi(X,t)$ is the rescaled random variable and $X:=x/\xi(t)$ is the coordinate rescaled by the correlation length $\xi(t):=\frac{2}{A}(\Gamma{}t)^{2/3}$.
The one-point distribution of $\chi(0,t)$ (and also for general $X$ if properly rescaled) is known to converge to the GOE Tracy-Widom distribution for the flat subclass, the GUE Tracy-Widom distribution for the circular subclass, and the Baik-Rains distribution for the stationary subclass.
This is expressed by $\chi(0,t) \tod \chi_1, \chi_2, \chi_0$.
In fact, equal-time multi-point distributions of $\chi(X,t)$ are also known and expressed as follows:
\begin{equation}
    \chi(X,t) \tod \begin{cases} \mathcal{A}_1(X) & \text{(flat)} \\ \mathcal{A}_2(X)-X^2 & \text{(circular)} \\ \mathcal{A}_{0}(X) & \text{(stationary)} \end{cases}
\end{equation}
in terms of certain stochastic processes called the Airy$_1$ process $\mathcal{A}_1(X)$ \cite{sasamoto_2005,borodin_2007a}, the Airy$_2$ process $\mathcal{A}_2(X)$ \cite{prahofer_2002}, and the Airy$_{\mathrm{stat}}$ process $\mathcal{A}_{0}(X)$ \cite{baik_2010}, though the Airy$_{\mathrm{stat}}$ process is essentially the double-sided Brownian motion plus a nontrivial random shift \cite{quastel_2014}.
Compared with equal-time properties, two-time correlation of $\chi(X,t)$ has been a challenging problem, but it has also been shown to have distinct properties among the three subclasses \cite{takeuchi_2012,ferrari_2016a,denardis_2017a} and some rigorous results are available now \cite{Johansson-PTRF2019,johansson_2021,liu_2021}.
Besides, for the stationary KPZ subclass, covariance of the height or the height gradient between points $(x,t)$ and $(0,0)$ in space-time has been studied and the exact solutions are available \cite{prahofer_2004}.

To characterize these multi-point correlations from data, it is convenient to define the rescaled height
\begin{align}
	q(X,t) := \frac{h(X\xi(t),t)-v_\infty t}{(\Gamma t)^{1/3}} \simeq \chi(X,t).  \label{eq:q_definition_sup}
\end{align}
With this, we define the spatial covariance
\begin{align}
    C_s(\Delta X,t) &:= \Cov{q(\Delta X,t)}{q(0,t)} \notag \\
    &\simeq \Cov{\chi(\Delta X,t)}{\chi(0,t)}
\end{align}
and the temporal covariance
\begin{align}
    C_t(t,t_0) &:= \Cov{q(0,t+t_0)}{q(0,t_0)} \notag \\
    &\simeq \Cov{\chi(0,t+t_0)}{\chi(0,t_0)}
\end{align}
where $\Cov{B}{C}:=\ave{BC}-\ave{B}\ave{C}$ is the covariance between $B$ and $C$.
Then, the spatial covariance $C_s(\Delta X,t)$ can be compared with the theoretical curves for the covariance of the Airy$_1$ and Airy$_2$ processes, which were made available by Bornemann \cite{Bornemann-MC2010} by numerical evaluation of the theoretical formula.
Concerning the temporal covariance, the theoretical functional form that can be compared with data is available only for the stationary subclass \cite{ferrari_2016a} to our knowledge.
However, the power laws describing the decay of the temporal covariance are available for all the three subclasses \cite{takeuchi_2012,ferrari_2016a,denardis_2017a}, as follows:
\begin{equation}
    C_t(t,t_0) \sim \begin{cases} (t/t_0)^{-1} & \text{(flat)}, \\ (t/t_0)^{-1/3} & \text{(circular)}, \\ (t/t_0)^{-1/3} & \text{(stationary)}. \end{cases}
\end{equation}

\begin{figure}[H]
	\centering
	\includegraphics[width=\linewidth]{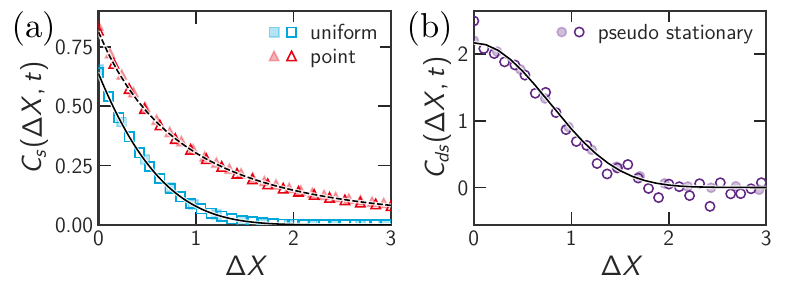}
	\caption{\label{fig:spatial_corr} 
		The spatial covariance for the uniform and point perturbations (a) and the two-point height-gradient covariance for the pseudo-stationary perturbations (b).
		(a) The spatial covariance $C_s(\Delta X,t)$ at $t=10^4$ (lighter color) and $t=10^5$ (darker color), for the uniform (turquoise squares) and point (red triangles) perturbations.
        The covariance of $\mathrm{Airy}_{1}$ and $\mathrm{Airy}_{2}$ processes is shown by the solid and dashed lines, respectively.
        (b) The two-point height-gradient covariance $C_{ds}(\Delta X,t)$ for the pseudo-stationary perturbations, at $t=10^4$ (lighter color) and $t=10^5$ (darker color), compared with the exact solution $g''(\Delta X)$ (solid line).
	}
\end{figure}

\begin{figure}[H]
	\centering
	\includegraphics[width=0.85\linewidth]{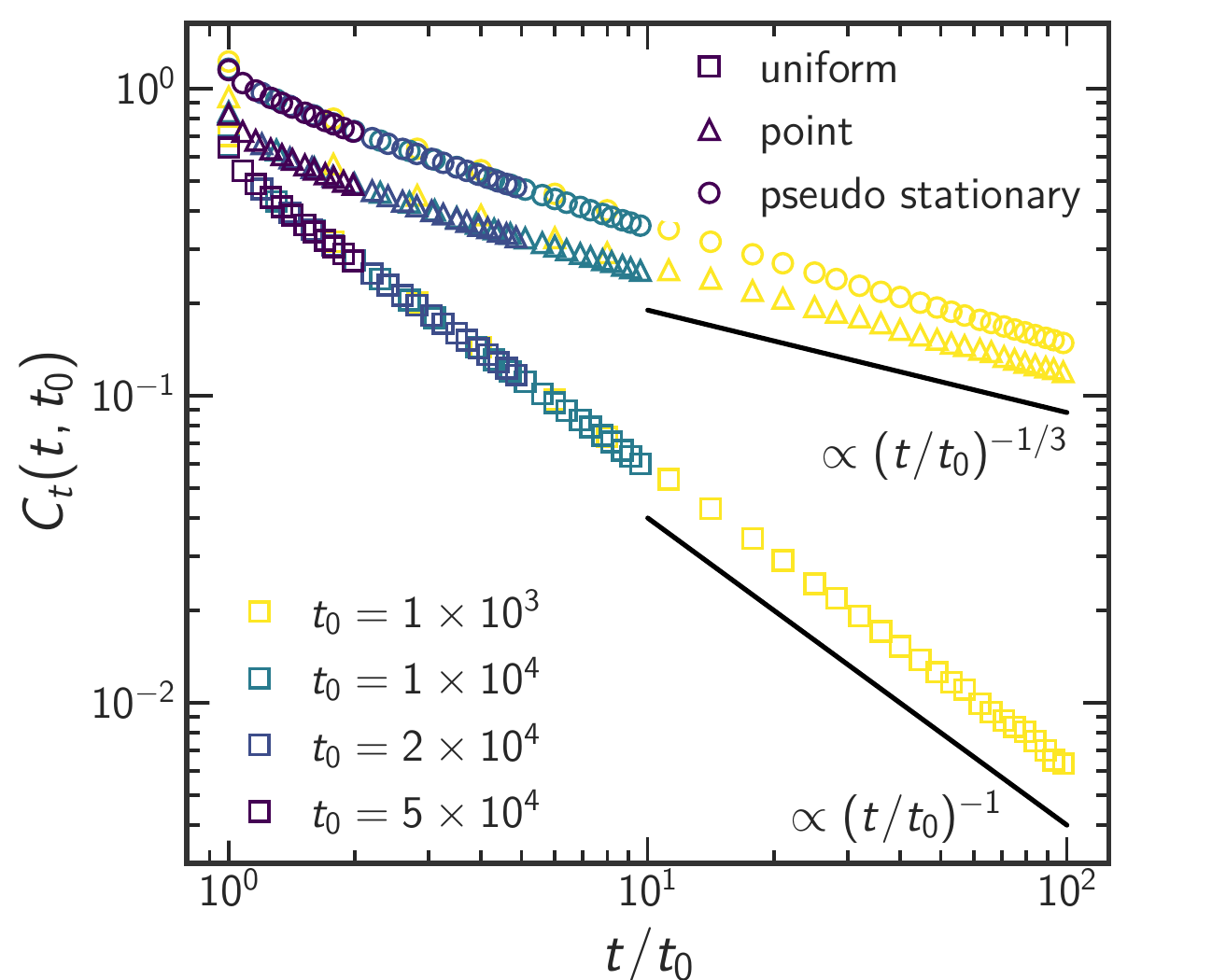}
	\caption{\label{fig:temporal_corr} 
		The temporal covariance $C_t(t,t_0)$. 
		The black lines are guides to the eyes.
	}
\end{figure}

For the uniform and point initial perturbations, we measured the spatial covariance $C_s(\Delta X,t)$ and found agreement with the Airy$_1$ and Airy$_2$ covariance, respectively [Fig.~\ref{fig:spatial_corr}(a)].
We also measured the temporal covariance $C_t(t,t_0)$ and found the power laws $(t/t_0)^{-1}$ and $(t/t_0)^{-1/3}$ for the uniform and point perturbations, respectively (Fig.~\ref{fig:temporal_corr}).
These constitute further evidence for the conclusion drawn in the article, that the perturbation interfaces for the uniform and point initial perturbations are described by the flat and circular KPZ subclasses, respectively.

For the pseudo-stationary initial perturbations, we measured the two-point height-gradient correlation function, defined by
\begin{equation}
    C_{ds}(\Delta X,t) := 2\ave{\frac{\partial q}{\partial X}(\Delta X,t)\frac{\partial q}{\partial X}(0,t)}.
\end{equation}
Here, we took both the ensemble average and the spatial average, using the fact that $h(x,t)-h_0(x)$ is statistically equivalent to $h(0,t)$.
Then we found agreement with the exact solution $g''(\cdot)$ by Pr\"ahofer and Spohn \cite{prahofer_2004} [Fig.~\ref{fig:spatial_corr}(b)].
We also measured the temporal covariance $C_t(t,t_0)$ and found the predicted power law $(t/t_0)^{-1/3}$ (Fig.~\ref{fig:temporal_corr}).
These demonstrate that the perturbation interfaces from the pseudo-stationary initial conditions are indeed described by the stationary KPZ subclass, as concluded in the article.

\section{Supplemental Text on Pseudo-Stationary Initial Conditions}
For pseudo-stationary initial conditions, we generated such a random walk $\tilde{h}_0(x)$ that starts from $\tilde{h}_0(0)=0$ and steps by $\tilde{h}_0(x+1)=\tilde{h}_0(x)+\sqrt{A}\mathcal{N}$, 
where $\mathcal{N}$ is a random variable generated from the standard normal distribution.
Then, to satisfy the periodic boundary condition, we set $h_0(x)=\tilde{h}_0(x)-\frac{x}{L}[\tilde{h}_0(L)-\tilde{h}_0(0)]$.

\newpage
\twocolumngrid
	
\section{Supplemental Figures}
\begin{figure}[H]
	\includegraphics[width=\linewidth]{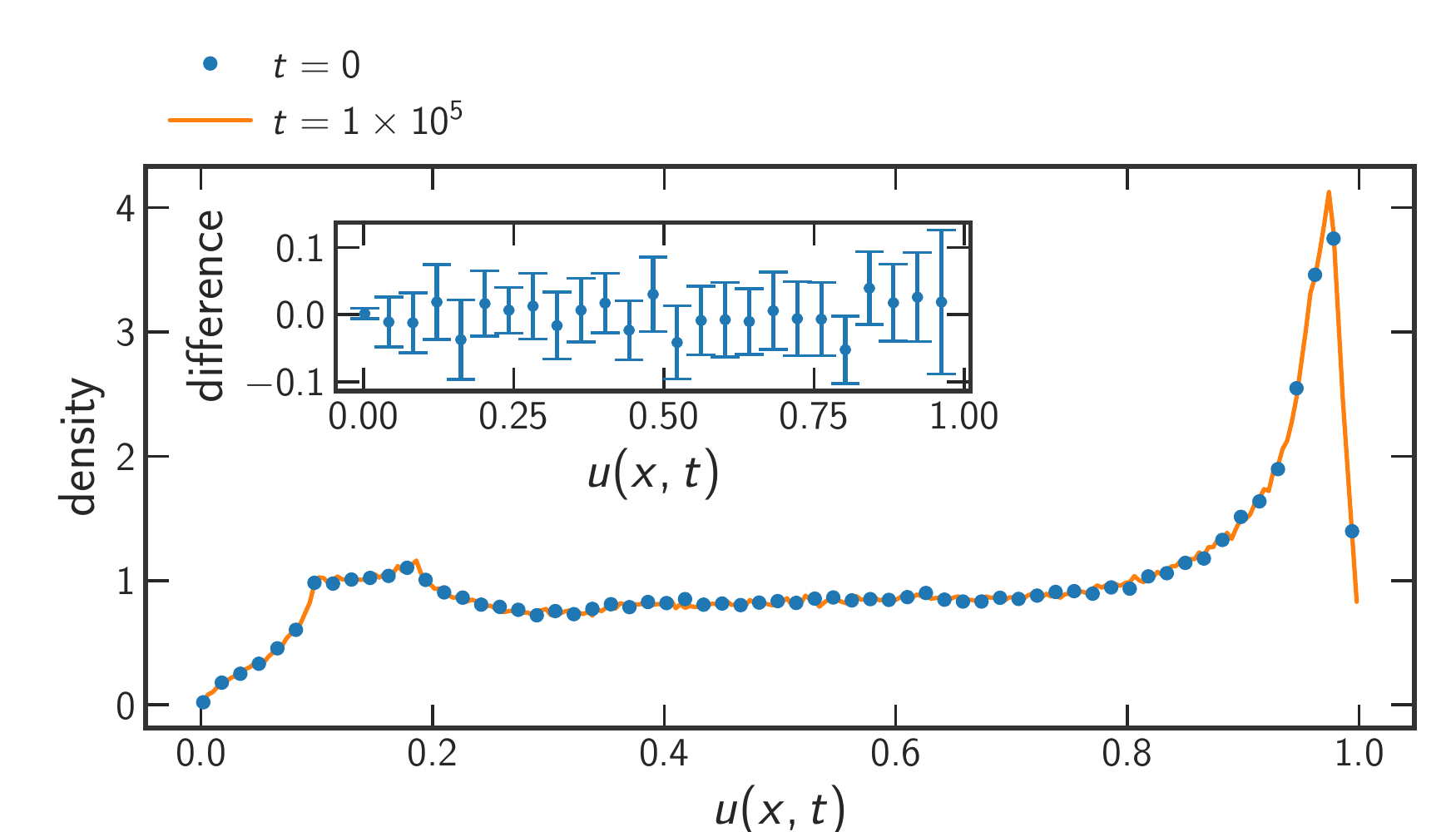}
	\caption{\label{fig:z_histogram} 
		The histogram of the dynamical variable $u(x,t)$, at $t=0$ and $t=1\times10^5$ with $T_\mathrm{warmup}=1000$ (see the main text).
		The inset shows the difference in the probability density, 
		where the error bars indicate the standard error.
	}
\end{figure}
	
\begin{figure}[H]
	\includegraphics[width=\linewidth]{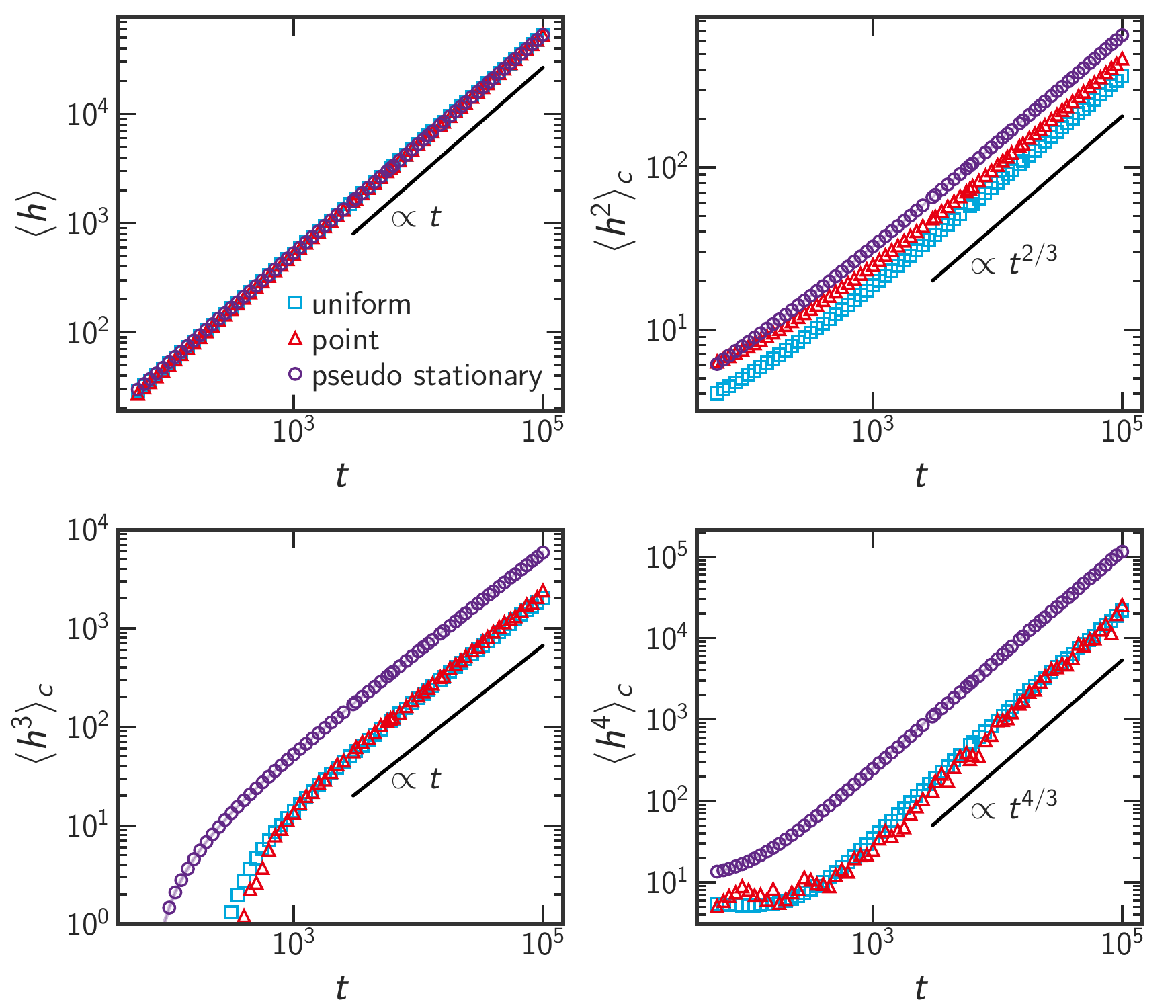}
	\caption{\label{fig:raw_cumulants} 
			The cumulants of the height, 		$\ave{h(0,t)^k}_c$, for the uniform, point, and pseudo-stationary initial perturbations			($k=1,2,3,4$). 
			The black solid lines are guides to the eyes.
	}
\end{figure}
	
\begin{figure}[H]
	\centering
	\includegraphics[width=0.85\linewidth]{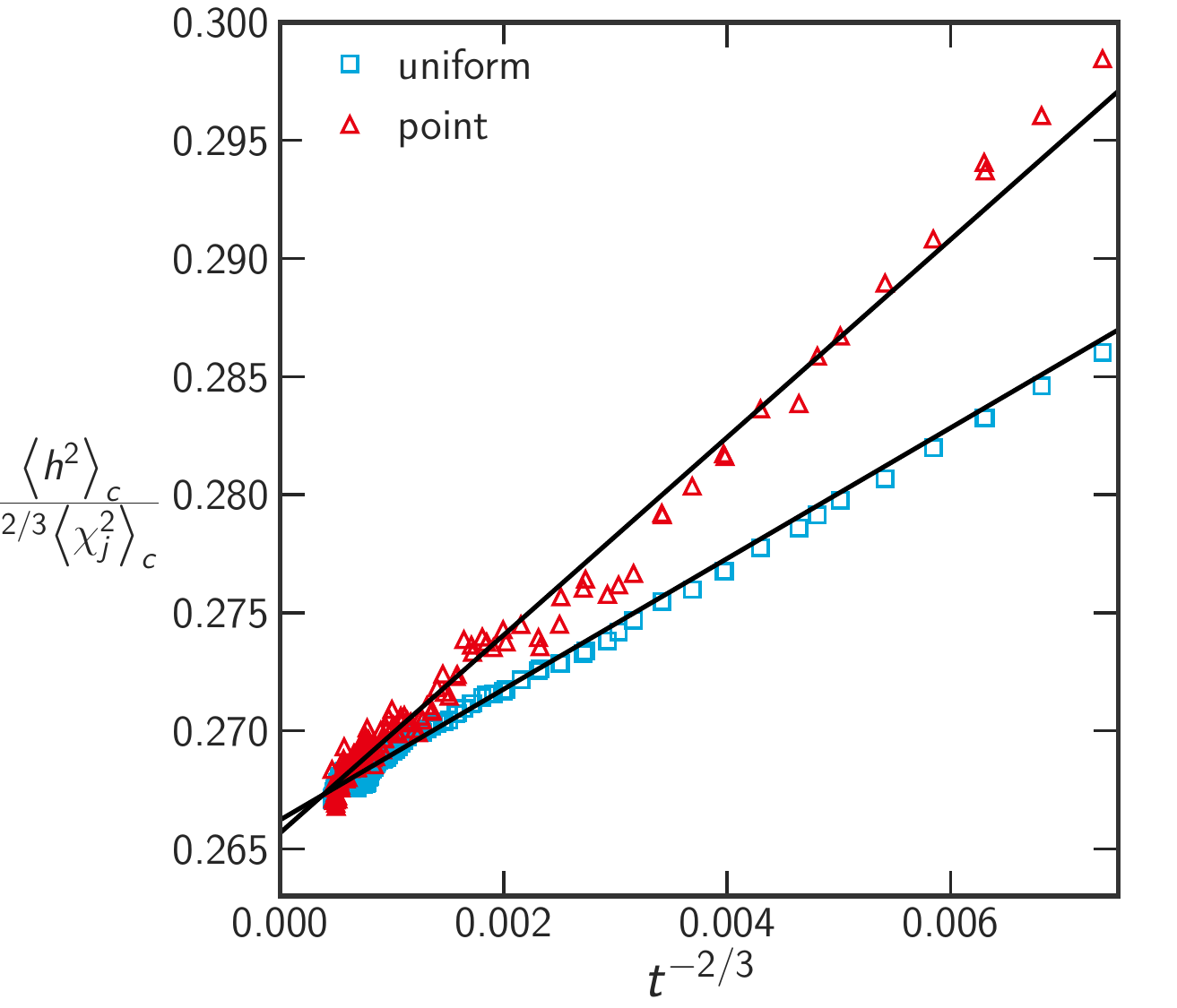}
	\caption{\label{fig:Gamma_estimation} 
		The variance of the height $\ave{h(0,t)^2}_c$ 
		rescaled by $t^{2/3}\ave{\chi_j^2}_c$, 
		where $j=1$ ($j=2$) for the uniform (point) initial perturbation.
		The black solid lines are linear fits used for the estimation of $\Gamma$.
	}
\end{figure}
\begin{figure}[H]
	\includegraphics[width=\linewidth]{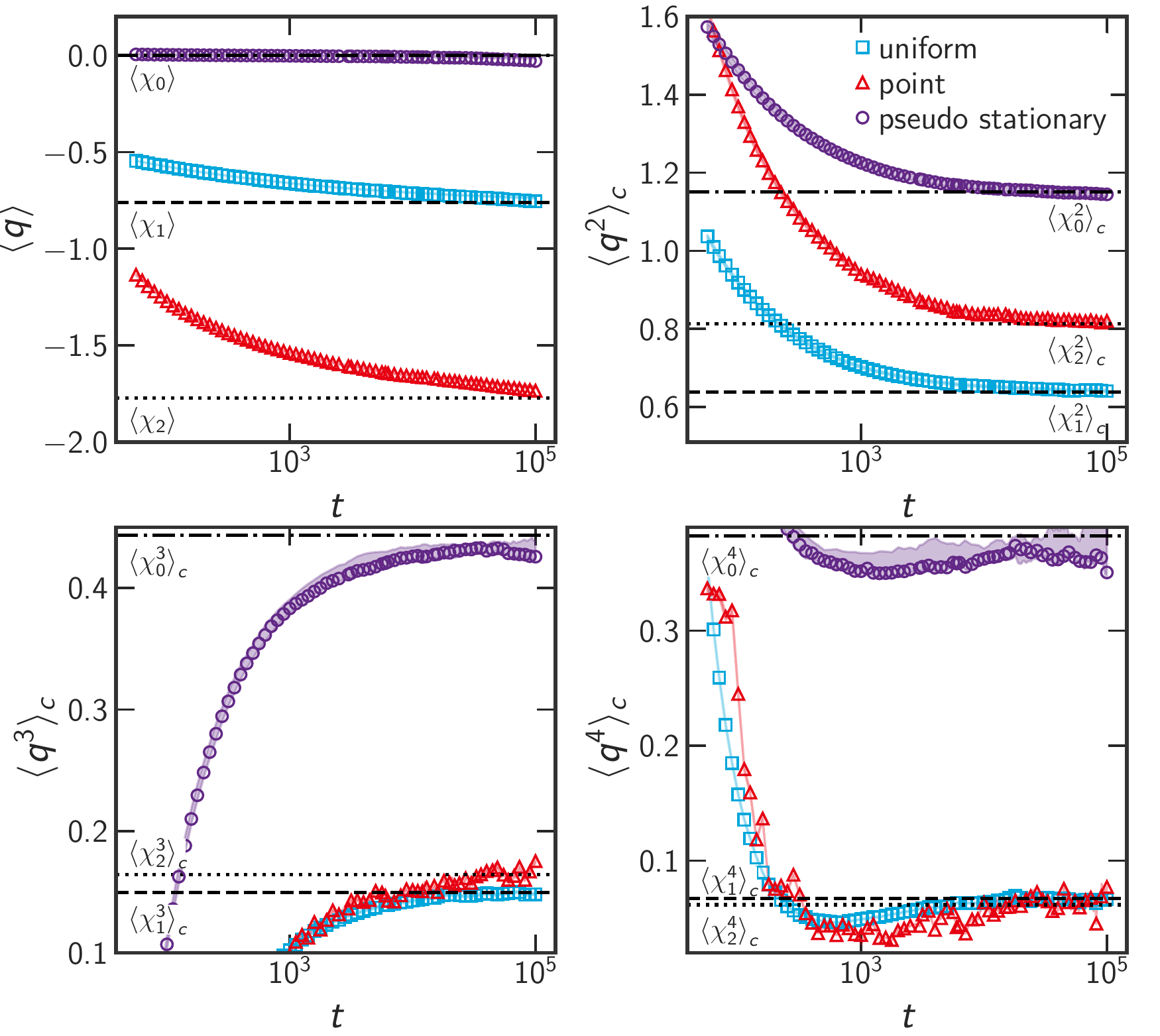}
	\caption{\label{fig:rescaled_cumulants} 
		The cumulants of the rescaled height $\ave{q(0,t)^k}_c$ ($k=1,2,3,4$). 
		The shaded area for the pseudo-stationary data indicates the range of the systematic uncertainty due to the parameter estimation.
		The values for $\chi_1$, $\chi_2$, and $\chi_0$ are shown by the dashed, dotted, and dash-dotted lines, respectively.
	}
\end{figure}

\section{Supplemental Tables}
\begin{table}[H]
	\centering
	\begin{tabular}{rrr}
		\hline
		\multicolumn{1}{c}{$L$} & \multicolumn{1}{c}{$T$} & \multicolumn{1}{c}{samples} \\
		\hline
		128 &  1448154 &    12000 \\
		256 & 40960000 &      480 \\
		512 & 11585237 &      480 \\
		1024 & 32768000 &      480 \\
		\hline
	\end{tabular}
	\caption{\label{tab:lambda_estimation_params} 
	The simulation parameters used to estimate $A\KPZlambda$. 
	Here, $T$ is the number of time steps.
}
\end{table}
\bibliography{citations}